\documentclass[a4paper,10pt]{article}

\usepackage[onehalfspacing]{setspace}
\usepackage{amsmath}
\usepackage{amssymb}
\usepackage{hyperref}
\usepackage{fontenc}
\usepackage{inputenc}
\usepackage{authblk}
\usepackage{color}
\usepackage{graphicx}
\usepackage{cite}
\usepackage{slashed}
\hypersetup{colorlinks=true,linkcolor=magenta,linktoc=page,citecolor=blue}
\usepackage{tikz}
\usepackage{multirow}
\usepackage{hhline}
\usepackage[top=1.5cm,bottom=2.5cm,right=2.54cm,left=2.54cm]{geometry}

\definecolor{c1}{rgb}{0.5,0,1}
\colorlet{aqua}{-red!75}

\newcommand{\be}{\begin{equation}}
\newcommand{\bea}{\begin{eqnarray}}
\newcommand{\eea}{\end{eqnarray}}
\newcommand{\ba}{\begin{array}}
\newcommand{\ea}{\end{array}}
\newcommand{\ee}{\end{equation}}
\usepackage{authblk}
\title{{\Large {\bf  Casimir Effect in Stochastic Semi-Classical Gravity}}}
\author[1]{M.Hossein. B.Khoshnevis}
\author[2]{Sadaf Ebadi}
\author[3]{M. Reza Lahooti}
\author[4]{Reza Pirmoradian}
\affil[1,2]{Department of Physics, Central Tehran Branch, Islamic Azad University (IAUCTB)}
\affil[3,4]{Ershad Damavand Institute for Higher Education (EDI)}

\begin{document}
	\maketitle
		

		
	
		

	\begin{abstract}
		
	This article aims to examine the Casimir effect in the framework of stochastic semi-classical gravity. We commence with the semi-classical Einstein-Langevin equation, which introduces a first-order correction to the semi-classical gravity theory. Subsequently, we analyze the alteration in the Casimir force caused by this type of correction. Our results demonstrate that the corrections exhibit significant sensitivity to both the distance between the two parallel plates and the order parameter of the weak field. This finding underscores the nuanced interplay between these factors and the overall behavior of the system, providing valuable insights into the underlying dynamics.
		
	\end{abstract}
	
\section{Introduction}

In 1948 a force of attraction between two parallel, neutral and conducting plates in a vacuum was predicted by Casimir  which was named Casimir force \cite{1,27}.  The Casimir effect itself arises from quantum fluctuations of the electromagnetic field between two uncharged, conducting plates placed very close to each other in a vacuum. These fluctuations cause an attractive force between the plates, which can be calculated using quantum field theory. The Casimir effect stands out as one of the most remarkable manifestations of non-trivial properties of the vacuum state in quantum field theory \cite{2,3} and this effect can be interpreted as the polarization of the vacuum by the boundary conditions or geometry. A straightforward computation of the Casimir pressure yields infinite sums, which therefore requires regularization \cite{4,5,6,7}. Several regulators may be applied, each yielding the same result; Green's function \cite{2,8}, stress tensor methods \cite{9, be,WGW, be2}, Krein space method \cite{Khosravi:2006tx,Krein, pej, fo}. The renormalization of the Casimir effect is a fascinating topic in quantum field theory and condensed matter physics. Formally, in the context of the Casimir effect, renormalization involves the following steps; Regularization: This step involves introducing a cutoff parameter to make the integrals finite. For example, one might impose a maximum frequency or momentum to avoid divergences. Renormalization: After regularization, the physical quantities are redefined to absorb the divergences. This is done by introducing counterterms into the Hamiltonian or Lagrangian of the system. Physical Predictions: The renormalized quantities are then used to make physical predictions that are finite and meaningful. Renormalization of the Casimir effect has several important applications and implications such as nanotechnology, where understanding the Casimir effect is crucial for designing nanoscale devices, as the forces can significantly affect the behavior of components at small scales. The Casimir effect can be generalized to other types of fields and materials, providing insights into phenomena such as van der Waals forces and surface interactions \cite{12}.\\
The relationship between the Casimir effect and weak gravity is an intriguing area of study that bridges quantum field theory and general relativity. In the regime of weak gravity, where gravitational fields are small perturbations to flat spacetime, the modifications to vacuum fluctuations become significant \cite{mil}. By studying the Casimir effect in curved spacetime, one can investigate how general relativity impacts quantum electrodynamics, this may provide insights into the quantum aspects of gravity. Theoretical predictions suggest that gravitational waves, ripples in spacetime caused by massive accelerating objects, could induce a Casimir-like force between closely spaced mirrors \cite{bor}. This potential gravitational Casimir effect could provide a novel experimental avenue for testing quantum gravity theories and searching for gravitons, the hypothetical quantum particles that mediate gravity \cite{dec}. Such studies highlight the interplay between quantum mechanics and general relativity and open new pathways for exploring fundamental physics (see also \cite{pir}).  Basically, this phenomenon could, in principle, occur for any quantized field so that if gravity has a quantum nature, then gravitational waves should also generate Casimir-like forces. In the regime of the semi-classical theory of gravity which describes a classical gravitational field coupled to a quantum matter field \cite{10}, vacuum polarization caused by the gravitational field is also referred to as the Casimir effect. The interaction between gravity and quantum fields was first explored in the context of cosmological back-reaction problems \cite{11,12new}. In the context of quantum field theory, zero-point field fluctuations contribute to the vacuum energy. Despite being fundamental, this energy remains unobservable in laboratory settings. Nevertheless, the vacuum energy density, more precisely the vacuum expectation value of the stress-energy tensor, is incorporated into Einstein's equation as it pertains to the vacuum state of the scalar field, thereby influencing various cosmological phenomena \cite{13,14} as
\begin{align}
\langle\hat T_{\mu\nu}\rangle_{vac}&=-\langle\rho\rangle_{vac}g_{\mu\nu}\nonumber\\
R_{\mu\nu}-\dfrac{1}{2}g_{\mu\nu}R+\Lambda g_{\mu\nu}&=8\pi G\langle T_{\mu\nu}\rangle_{vac}, \label{vac}
\end{align} $R_{\mu\nu}$, $R$, $\Lambda$ and $ G$ are, respectively, Ricci tensor, Ricci scalar, cosmological constant and Newton's gravitational constant. The classical metric's equation of motion is the semi-classical Einstein's equation which describes the back reaction of the matter field on the spacetime; it is a generalized form of the Einstein's equation where the energy source is supposed to be the expectation value in some quantum state of the matter stress energy tensor operator. The lowest order quantum stress energy fluctuations of the matter field as a source of classical stochastic fluctuations of the gravitational field is described by the semi-classical Einstein-Langevin equation \cite{15}.

On the other hand, stochastic gravity deals with the effects of quantum fluctuations in spacetime, particularly focusing on the gravitational back-reaction from quantum fields, which can result in phenomena such as the generation of gravitational waves or the understanding of black hole thermodynamics. Stochastic gravity starts with quantum field theory in curved spacetime \cite{16,17,18,19,20}. In fact the physical observable that measures the stress-energy fluctuations may be considered as an advantage of the derivation of Einstein-Langevin equation and of the stochastic gravity theory \cite{21}. In the context of stochastic gravity, Casimir force might be interpreted as part of the broader discussion of quantum field theory in curved spacetime, where the random nature of quantum fluctuations interacts with gravitational fields.  The interplay between quantum forces, such as the Casimir force, and gravity in stochastic gravity frameworks provides important insights into the behavior of the quantum vacuum and its potential influence on large-scale cosmic structures.  The aim of this paper is to investigate the Casimir force in the linear form in the stochastic gravity framework. 

The paper is structured as follows:  Sections 2 and 3 review the mathematical foundations underpinning this study, focusing on semi-classical gravity and the Casimir force. Section 4 applies these results to explore the Casimir force within the weak field limit of gravity. Finally, the paper wraps up with a discussion of the conclusions derived from the investigation.

\section{Semi-classical Gravity: Mathematical Framework }
\label{sect:basics}
Semi-classical gravity is a theoretical approach that unifies classical general relativity with quantum field theory. In this approach, gravity is treated classically through Einstein's field equations, while other fields, such as the electromagnetic field, are quantized. The Casimir effect, which arises from quantum fluctuations of the electromagnetic field between two uncharged, closely spaced conducting plates, can be studied within this context. By examining the Casimir effect in semi-classical gravity, one can explore how quantum field fluctuations influence and are influenced by the curvature of spacetime. This interplay offers insights into the quantum aspects of gravity and the effects of quantum vacuum energy on gravitational systems. In fact  Semi-classical gravity describes the interaction of gravitational field, which is assumed to be a classical field, with quantum matter fields \cite{22}. The theory at least for the linear matter field, is mathematically well-defined and fairly well-understood  \cite{19}.  The classical action for a massive real scalar $\phi(x)$ is given by
\begin{equation}
S_m[g,\phi]=-\dfrac{1}{2}\int d^{4}x\sqrt{-g}[g^{\mu\nu}\nabla_{\mu}\phi\nabla_{\nu}\phi+(m^{2}+\xi R )\phi^{2}]
\end{equation}
where $\xi$ is called conformal coupling. The Klein-Gordon equation  can be written for a real scalar field as 
\begin{equation} \label{4}
(\square-m^{2}-\xi R)\hat \phi =0,
\end{equation}
note that $\square=\nabla_\mu \nabla^{\mu}$. The stress-energy tensor is given by
\begin{equation}
T_{\mu \nu }\equiv \dfrac {-2}{\sqrt {-detg}}\dfrac {\delta S_m }{\delta g^{\mu \nu }},
\end{equation}
 which leads to 
\begin{equation}
T_{\mu\nu}[g,\phi]=\nabla_\mu \phi \nabla_\nu \phi-\dfrac{1}{2}g_{\mu\nu}(\nabla^{\rho}\phi\nabla_{\rho}\phi+m^{2}\phi^{2})+\xi(g_{\mu\nu}\square-\nabla_\mu\nabla_{\nu}+G_{\mu\nu})\phi^{2},
\end{equation}
  $G_{\mu\nu}$ is the Einstein tensor. In the semi-classical theory the gravitational field equation will not be the classical equation instead, the source is supposed to be the quantum expectation value $\langle\hat T_{uv}\rangle[g] $.  The semi-classical Einstein equation for the metric $g_{uv}$ is given by
\begin{equation}\label{3}
\dfrac{1}{8\pi G}(G_{\mu\nu}[g]+\Lambda g_{\mu\nu})-2(\alpha A_{\mu\nu}+\beta B_{\mu\nu})[g]=\langle\hat T^{R}_{\mu\nu}\rangle[g],
\end{equation}
 $\langle\hat T^{R}_{\mu\nu}\rangle[g]$ is the expectation value of $\hat T^{R}_{\mu\nu}$ defined by
\begin{equation}\label{8}
\hat T^{R}_{\mu\nu}[g](x)=\hat T_{\mu\nu}[g](x)+ F^{C}_{\mu\nu}[g](x)\hat I,
\end{equation}
where $\hat I$ is the identity operator and $F^{C}_{\mu\nu}[g]$ are some symmetric tensor counter terms that depend on the regulator and local function of the metric. Left hand side of the equation \eqref{3} may be derived from the following gravitational action
\begin{equation}\label{9}
S_g[g]=\int d^{4}x\sqrt{-g}[\dfrac{1}{16\pi G}(R-2\Lambda)+\alpha C_{\mu\nu\rho\sigma}C^{\mu\nu\rho\sigma}+\beta R^{2}],
\end{equation}
 where $C_{\mu\nu\rho\sigma}$ is the Weyl tensor. $A_{\mu\nu}$ and $B_{\mu\nu}$ are tensors that are derived from derivatives of the functional with respect to the metric  of the quadratic terms in the curvature in \eqref{9} which are explicitly given by
\begin{align}
A_{\mu\nu}&=\dfrac{1}{2}g_{\mu\nu}C_{\tau\epsilon\rho\sigma}C^{\tau\epsilon\rho\sigma}
-2R_{\mu\rho\sigma\tau}R_{\nu}^{\rho\sigma\tau}+4R_{\mu\rho}R^{\rho}_{\nu}-\dfrac{2}{3}RR_{\mu\nu}-2\square R_{\mu\nu}+\dfrac{2}{3}\nabla_{\mu}\nabla_{\nu}R+\dfrac{1}{3}g_{\mu\nu}\square R,\nonumber\\
B_{\mu\nu}&=\dfrac{1}{2}g_{\mu\nu}R^{2}-2RR_{\mu\nu}+2\nabla_\mu\nabla_\nu R-2g_{\mu\nu}\square R,
\end{align}
 $R_{\mu\nu\rho\sigma}$ and $R_{\mu\nu}$ are Riemann and Ricci tensors respectively.\\

Stochastic gravity is an advanced framework that blends quantum field theory with general relativity to investigate the impact of quantum fluctuations on gravitational phenomena. Within this framework, zero-point field fluctuations contribute to the vacuum energy, which remains unobservable in laboratory settings. By rewriting the equation \eqref{vac} as the form of the Einstein-Langevin equation  which is central to stochastic gravity given by
\[ G_{\mu\nu} + \Lambda g_{\mu\nu} = 8\pi G \langle T_{\mu\nu} \rangle + \sqrt{8\pi G} \int d^4x' \, N_{\mu\nu}(x,x') \]
where \( \langle T_{\mu\nu} \rangle \) is the expectation value of the stress-energy tensor. The term \( N_{\mu\nu}(x,x') \) is the noise kernel, defined as:
\be N_{\mu\nu}(x,x') = \langle \delta T_{\mu\nu}(x) \delta T_{\mu\nu}(x') \rangle \ee
which captures the vacuum fluctuations in the stress-energy tensor. Metric perturbations around Minkowski spacetime can be represented as \( h_{\mu\nu} = g_{\mu\nu} - \eta_{\mu\nu} \). The fluctuation-dissipation relation, 
\be \langle \delta T_{\mu\nu}(x) \delta T_{\mu\nu}(x') \rangle = \frac{1}{2} \left( \frac{\delta \langle T_{\mu\nu} \rangle}{\delta g^{\mu\nu}(x)} \right) \delta^4(x - x'), \ee
provides a connection between the fluctuations in the stress-energy tensor and the dissipative dynamics of metric perturbations. These are the key formulas that illustrate how stochastic gravity integrates quantum effects into classical gravitational theory, providing a robust framework for exploring quantum fluctuations' influence on gravity \cite{23,24}. Stochastic semi-classical gravity aims to extend the semi-classical theory by accounting for these fluctuations in a self-consistent way. We use this approach to study the Casimir effect.

\section{Casimir Effect in Minkowski Space-time}

  The Casimir effect has been studied for different fields in different background geometries \cite{5,6,17,25,27}. For two large, perfectly conducting plates with surface area $A$, separated by a distance $L$, where $\sqrt{A}>>L$, allowing us to neglect edge contributions. Plates are in $x-y$ plane at $Z=0$ and $Z=L$. The attractive force per unit area, i.e. the pressure between two infinitely large, neutral parallel planes made of an ideal metal at zero temperature with Dirichlet boundary condition for a massless minimally coupled field in the Minkowski space-time, where the Casimir energy per unit area is $\mathcal{E}_C=-\dfrac{\pi^{2}}{720L^{3}}$, is given by \cite{28}
 \begin{equation}
P(L)_{Mink}=\dfrac{-\pi^{2}}{240L^{4}}.
\end{equation}
For a massless conformally coupled scalar field in the flat space-time, the stress-energy-momentum tensor in Minkowski space-time in the massless conformal coupling is given by 
\begin{equation} 
T_{\mu\nu}=\dfrac{2}{3} \partial_\mu \phi \partial_\nu \phi+\dfrac{1}{6} \eta_{\mu\nu}\partial_\alpha \phi \partial^{\alpha}\phi -\dfrac{1}{3}\phi\partial_\mu  \partial_\nu \phi.
\end{equation}
 The vacuum energy density of the scalar field in the presence of boundaries is the mean value of the 00-component of the energy-momentum tensor in the vacuum state. In the conformally coupled case it is given by
\begin{equation}
\langle0|\hat T_{00}|0\rangle=\dfrac{5}{6}\langle0|\partial_{0}\hat \phi \partial_0 \hat \phi|0\rangle-\dfrac{1}{6}\langle0|\partial_i \hat \phi \partial_i \hat \phi|0\rangle-\dfrac{1}{3}\langle0|\hat \phi \partial^{2}_0 \hat \phi|0\rangle
\end{equation}
 it can be easily shown that
\begin{equation}
\partial_i (\hat \phi \partial_i \hat \phi)=-\partial_i \hat \phi \partial_i \hat \phi
\end{equation}
thus, the expectation value of the energy-momentum tensor (00-component) leads to
\begin{equation}\label{29}
\langle0|\hat T_{00}|0\rangle=\bigg[\dfrac{5}{6}\partial^{x}_0\partial^{y}_0+\dfrac{1}{6}\partial^{x}_i\partial^{y}_i\bigg]\langle0|\hat \phi(x)\hat \phi(y)|0\rangle.
\end{equation}
To compute the above relation we use the Feynman propagator is defined by 
\begin{align} 
iG_F(x,x^{'})&=\langle0|T(\hat\phi(x)\hat\phi(x^{'})|0\rangle\nonumber\\
&=\theta(t-t^{'})G^{+}(x,x^{'})+\theta(t^{'}-t)G^{-}(x,x^{'})
\end{align}
where the Wightman functions $G^{\pm}$ are defined as follows
\begin{align}
G^{+}(x,x^{'})&=\langle0|\hat\phi(x)\hat\phi(x^{'})|0\rangle,\nonumber\\
G^{-}(x,x^{'})&=\langle0|\hat\phi(x^{'})\hat\phi(x)|0\rangle.
\end{align}
$\theta(t)$ is the Heaviside step function, it is then straightforward to verify that 
\begin{align}
(\partial^{2}_t -\partial^{2}_i)G_F(x,x^{'})=i\delta^{4}(x-x^{'}).
\end{align}
 The general form of all Green's functions in which we use the mode decomposition of $\phi$ can be demonstrated as
\begin{align}
\mathcal{G}(x,x^{'})=\int \dfrac{d^{n}k}{(2\pi)^{n}}\dfrac{exp[i\vec{k}.(\vec{x}-\vec{x}^{'})-ik^{0}(t-t^{'})]}{(k^{0})^{2}-|\vec{k}|^{2}-m^{2}}.
\end{align}
where for Feynman propagator, we would shift the poles at $k^{0}=\pm(|k|^{2}+m^{2})^{1\over{2}}$ off the real axis by the replacement $m^{2}\rightarrow m^{2}-i\epsilon$ to recover $G_F(x,x^{'})$. As a consequence, in our case, it becomes
\begin{equation}
G_F(x,x^{'})=\dfrac{1}{4\pi^{2}}\dfrac{1}{(x-x^{'})^{2}}.
\end{equation}
The Casimir energy density arises from the quantum fluctuations of the electromagnetic field in the vacuum, influenced by the boundary conditions imposed by conducting surfaces or materials. On an interval, such as between two parallel conducting plates, these boundary conditions restrict the possible modes of the field, leading to a modification of the zero-point energy. The resulting Casimir energy density, which can be attractive or repulsive depending on the configuration, manifests as a measurable force between the plates, often referred to as the Casimir effect. This energy density is inversely proportional to the fourth power of the distance between the plates, highlighting its significance at microscopic scales. To calculate the Casimir force from the energy density, integrate the energy density over the spatial interval to obtain the total Casimir energy. Then, derive the Casimir force by taking the negative gradient (derivative) of the total Casimir energy with respect to the distance between the plates. As a result, by using the Eq.\eqref{29}, the Casimir energy density on the interval $0<z< L$ can be obtained  using the local version of the bellow equation
\begin{align} \label{eq22}
\epsilon=\langle0|\hat T_{00}|0\rangle^{L}-\langle0|\hat T_{00}|0\rangle^{\infty}=&-\dfrac{1}{32\pi^{2}}\sum_{n\neq0}\dfrac{1}{(nL)^{4}}\nonumber\\
=&-\dfrac{1}{16\pi^{2}L^{4}}\sum^{\infty}_{n=1}\dfrac{1}{n^{4}}=-\dfrac{1}{16\pi^{2}L^{4}}\zeta(4).
\end{align}
where, the zeta function is given by
\begin{equation}
\zeta(4)=\sum^{\infty}_{n=1}\dfrac{1}{n^{4}}=\dfrac{\pi^{4}}{90}.
\end{equation}
Thus, one obtains
\begin{equation}
\epsilon=\langle0|\hat T_{00}|0\rangle^{L}-\langle0|\hat T_{00}|0\rangle^{\infty}=-\dfrac{\pi^{2}}{1440L^{4}}.
\end{equation}
\begin{figure}[tbp]
	\centering 
	\includegraphics[width=0.7\textwidth,origin=c]{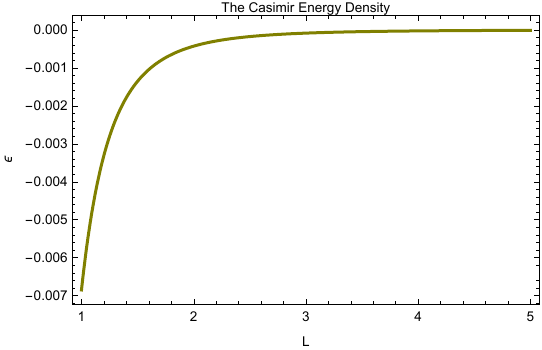}
	\caption{\label{fig1}Casimir energy density as  function of distance L }
\end{figure}
Now, we consider a factor of 2 to take into account the two polarizations of the electromagnetic field. Hence, the Casimir energy per unit area becomes as follows
\begin{align}\label{eq25}
\mathcal{E}_{C}=-\dfrac{\pi^{2}}{720L^{3}},
\end{align}
and the Casimir force per unit area is given by
\begin{equation} \label{eq26}
P(L)=-\dfrac{\pi^{2}}{240L^{4}}.
\end{equation}

\begin{figure}[tbp]
\centering 
\includegraphics[width=0.7\textwidth,origin=c]{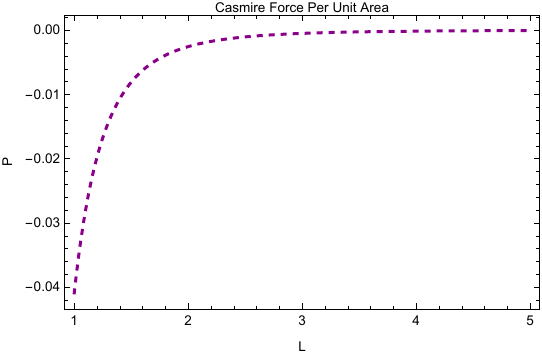}
\caption{\label{fig2}Casimir force per unit area as  function of distance L }
\end{figure}

\begin{figure}[tbp]
\centering 
\includegraphics[width=0.7\textwidth,origin=c]{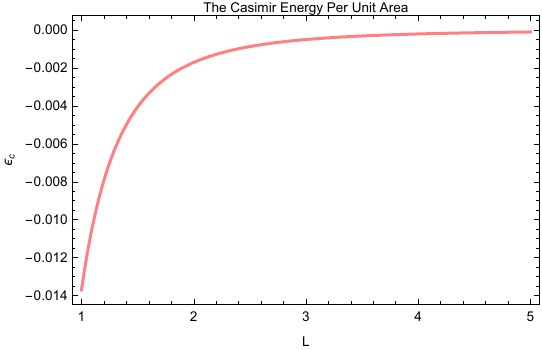}
\caption{\label{fig3} Casimir energy per unit area as  function of distance L}
\end{figure}
 
\section{ Gravity and Casimir Force: Weak Field Limit }

As mentioned, the Casimir effect, a quantum phenomenon resulting from vacuum fluctuations of the electromagnetic field, can be significantly influenced by the curvature of spacetime. In the context of curved space, the boundaries or geometries that constrain the field, such as conducting plates, experience modified quantum fluctuations due to the presence of gravitational fields. This leads to changes in the Casimir force, which can be either enhanced or reduced depending on the curvature. Studying the Casimir effect in curved space offers a unique window into understanding how quantum field theory and general relativity interact, providing insights into the quantum nature of gravity and the behavior of quantum fields in non-flat spacetime geometries. This intersection of quantum mechanics and general relativity helps advance our knowledge of fundamental physics and could potentially inform theories of quantum gravity. In a general curved space-time, we have shown $\nabla_\mu T^{\mu\nu}=0$, and in a space-time having special symmetries or isometries ($\varsigma$ generates an isometry $\mathcal{L}_\varsigma g_{\mu\nu}=\nabla_\mu \varsigma_\nu+\nabla_\nu \varsigma_\mu=0$) as implied by the existence of one or more killing vector fields $\varsigma^{\mu}$, we have $\nabla_\mu(T^{\mu\nu}\varsigma_\nu)=0$ and $\mathcal{P}_\varsigma\equiv \int dV_x T^{0}_\nu \varsigma^{\nu}$ is constant \cite{29}, where $dV_x=d^{n-1}x|g|^{1\over{2}}$.\\
 Stochastic gravity, on the other hand, extends the integration of quantum field theory and general relativity by incorporating quantum fluctuations into the gravitational field equations through the Einstein-Langevin equation. This advanced framework considers the noise kernel, representing stress-energy tensor fluctuations, to explore gravitational phenomena such as spacetime stability, structure formation, and the backreaction of quantum effects on black holes. Combining these two concepts, the Casimir effect demonstrates the observable consequences of quantum fluctuations, while stochastic gravity provides a theoretical foundation for understanding their impact on the curvature of spacetime. Here, let us consider a Casimir apparatus of parallel plates separated by a distance $L$, with transverse dimensions $a>>L$. Brown and Maclay \cite{30} showed that, in the z-direction (the direction of gravity), the vacuum expectation value stress-energy of a conformally invariant scalar field is given by 
\begin{align}
\langle T^{\mu\nu}\rangle=\dfrac{\mathcal{E}_C}{L}diag(1,-1,1,3)
\end{align}								
The Cartesian coordinates associated with the Casimir apparatus are ($\tilde \xi,\tilde \eta, \tilde \chi$) where $\tilde \xi$ is normal to the plate, while $\tilde \eta$ and $\tilde \chi$ are parallel to the plate. The apparatus is oriented at an angle $\alpha$ with respect to the direction of gravity. Where
\begin{equation}
z=\tilde \xi \cos \alpha+ \tilde \eta \sin \alpha,\hspace*{0.4cm} y=\tilde \eta \cos \alpha -\tilde \xi \sin \alpha, \hspace*{0.4cm} x=\tilde \chi.
\end{equation}
Let the center of the apparatus be located at ($\tilde\xi=\tilde\xi_0,\tilde{\eta}=0,\tilde \chi=0$). Gauge transformation of $h_{\mu\nu}$ for stochastic gravity obeys the following equation 
\begin{equation} 
h_{\mu\nu}\rightarrow h_{\mu\nu}+\partial_\mu \varsigma_\nu +\partial_\nu \varsigma_\mu.
\end{equation}
Thus, the action changes as follows  
\begin{align}
\Delta W=-2\int(dx)\varsigma_\nu \partial_\mu T^{\mu\nu}
\end{align}
To study the effect of gravity on the Casimir effect, we use Fermi coordinates. A Fermi coordinates system is the general relativistic generalization of an inertial coordinate frame. The "constant field metric" is simply the Fermi coordinate metric for gravitating body \cite{31}
\begin{align}
ds^{2}=-(1+2gz)dt^{2}+d\bf{r}^{'2}.
\end{align}
We compute the additional gravitational energy, in terms of the gauge field $\varsigma_\mu$, which carries us from isotropic coordinates to Fermi coordinates $h^{F}_{\mu\nu}=h^{I}_{\mu\nu}+\partial_\mu\varsigma_\nu+\partial_\nu\varsigma_\mu$. Therefore, we have
\begin{equation}
h^{I}_{00}=-gz,\hspace*{0.4cm}h^{I}_{ij}=-gz\delta_{ij},\hspace*{0.4cm}h^{F}_{00}=-gz,\hspace*{0.4cm}h^{F}_{ij}=0.
\end{equation}
With the proper definition of gauge field for the change in the energy obtained from $\Delta W$, finally we get
\begin{align}
\begin{split}
\Delta E_g&= \dfrac{6\mathcal{E}_C}{L}\int^{a\over{2}}_{-{a\over{2}}} d\tilde\eta \int^{a\over{2}}_{-{a\over{2}}} d\tilde \chi \dfrac{1}{4}g\cos\alpha(-2\tilde\xi_0L)-\dfrac{2\mathcal{E}_C}{L}\int^{\tilde\xi_0+{L\over{2}}}_{\tilde\xi_0-{L\over{2}}}d\tilde\xi  \int^{a\over{2}}_{-{a\over{2}}}d\tilde\chi\dfrac{1}{2}g\cos\alpha(-a)\tilde\xi\nonumber\\
&-\dfrac{2\mathcal{E}_C}{L}\int^{\tilde\xi_0+{L\over{2}}}_{\tilde\xi_0-{L\over{2}}}d\tilde\xi\int^{a\over{2}}_{-{a\over{2}}} d\tilde\eta\dfrac{1}{2}g(\tilde\xi \cos\alpha+\tilde\eta\sin\alpha)(-a)=-Ag\mathcal{E}_C\tilde\xi_0\cos\alpha=-Ag\mathcal{E}_Cz_0
\end{split}
\end{align}
Also we can write 
\begin{equation}\label{eq33}
-\dfrac{\delta\Delta E_g}{A\delta z_0}=\dfrac{\Delta F}{A}=g\mathcal{E}_C.
\end{equation}
When we add it to the isotropic force ($F^{I}=-2g\mathcal{E}_C$), we obtain the Fermi force as follows \cite{31,32,pir,34}
\begin{equation}
\dfrac{F^{I}+\Delta F}{A}=-2g\mathcal{E}_C+g\mathcal{E}_C.
\end{equation}
At the end, we get 
\begin{equation}
-g\mathcal{E}_C=\dfrac{F^{F}}{A}.
\end{equation}

\begin{figure}[tbp]
\centering 
\includegraphics[width=0.65\textwidth,origin=c]{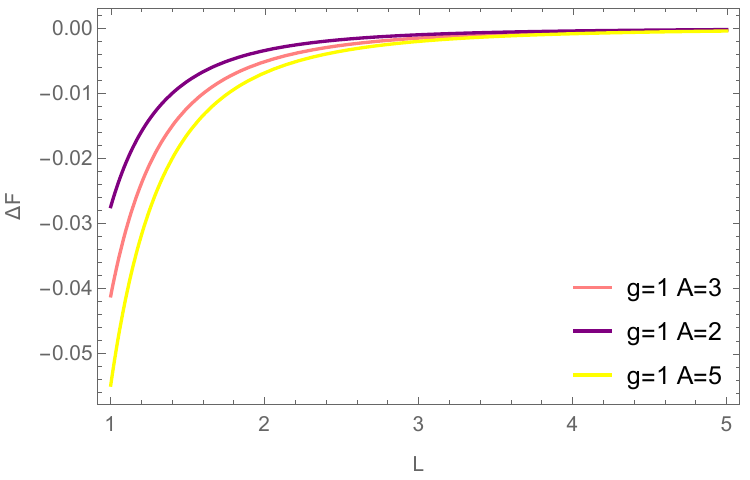}
\caption{\label{fig4}change in Force as a function of distance L for different area A }
\end{figure}

\begin{figure}[tbp]
\centering 
\includegraphics[width=0.65\textwidth,origin=c]{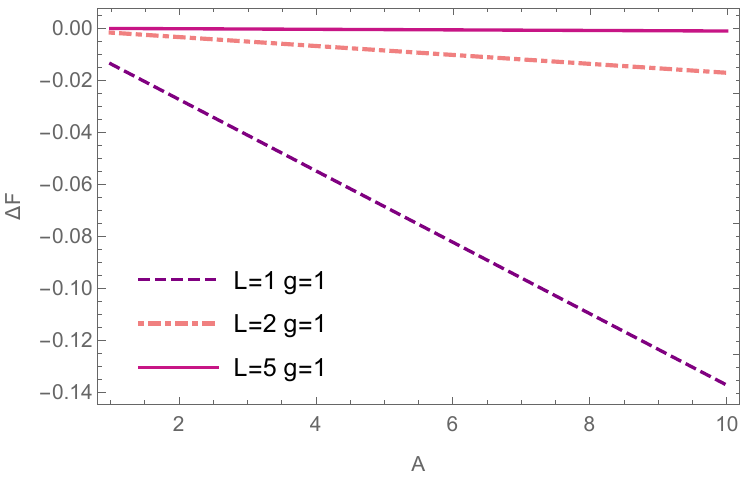}
\caption{\label{fig5}change in Force as a function of area A  for different distance L}
\end{figure}
\begin{figure}[tbp]
	\centering 
	\includegraphics[width=1.0\textwidth,origin=c]{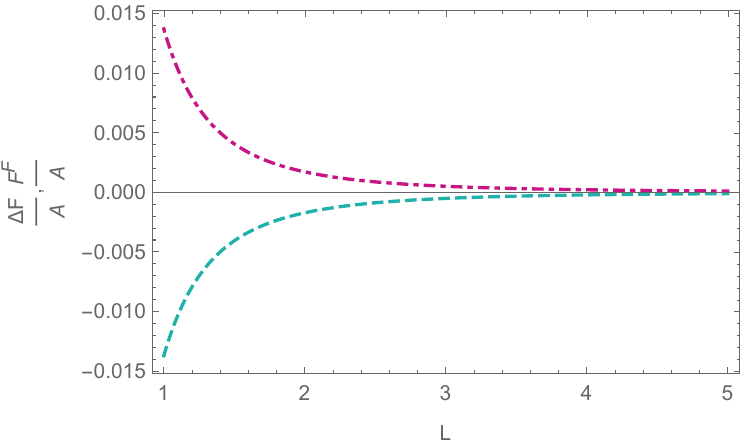}
	\caption{\label{fig6}Fermi force and change in force per unit area as  function of distance L }
\end{figure}
\section*{Concluding Remarks}
 By incorporating the Einstein-Langevin equation to account for stochastic fluctuations, we have derived first-order corrections to the semi-classical gravity framework. Our results demonstrate that these corrections are highly sensitive to both the distance between parallel plates and the order parameter of the weak field. This sensitivity underscores the intricate interplay between quantum fluctuations and gravitational influences, revealing that even subtle variations in these parameters can substantially alter the Casimir force. These findings not only enhance our understanding of the Casimir effect within the context of stochastic gravity but also highlight the broader implications of quantum field fluctuations on macroscopic gravitational phenomena. Further exploration of this interplay could potentially unlock new avenues in quantum gravity research and its applications in modern physics. 
 
 In conclusion, our investigation into the effects of weak field stochastic gravity on the Casimir force has yielded results which illustrated in the figures.  In Figure \ref{fig1}, Equation \ref{eq22} is utilized to plot the Casimir energy density as a function of the distance between the plates. Figure \ref{fig2} illustrates the Casimir force per unit area, derived from Equation \ref{eq26}, which is inversely proportional to the fourth power of the distance between the plates. Furthermore, as indicated by Equation \ref{eq25}, the Casimir force is shown to be inversely proportional to the cube of the distance between the plates, as depicted in Figure \ref{fig3}. Figure \ref{fig4} employs Equation \ref{eq33} to plot the change in force for different areas \(A\), while Figure \ref{fig5} shows the change in force as a function of area \(A\) for varying distances. Finally, Figure \ref{fig6} presents the change in force per unit area (light-green-sea dashed) and the Fermi force per unit area (violet-red dot-dashed) as functions of the distance \(L\).

\subsection*{Acknowledgments}

The authors would like to thank M. Amraji for his corporation and
interest in this work.

\end{document}